\documentclass[twocolumn,prl,showpacs,superscriptaddress,preprintnumbers,amsmath,amssymb,floatfix]{revtex4}
\usepackage{graphicx}
\usepackage{dcolumn}
\usepackage{bm}

\usepackage{bm}

\begin{document}

\title{Spin blockade, orbital occupation and charge ordering in La$_{1.5}$Sr$_{0.5}$CoO$_{4}$}

\author{C.~F.~Chang}
  \affiliation{II. Physikalisches Institut, Universit\"{a}t zu K\"{o}ln, Z\"{u}lpicher Str. 77, 50937 K\"{o}ln, Germany}
\author{Z.~Hu}
  \affiliation{II. Physikalisches Institut, Universit\"{a}t zu K\"{o}ln, Z\"{u}lpicher Str. 77, 50937 K\"{o}ln, Germany}
\author{Hua~Wu}
  \affiliation{II. Physikalisches Institut, Universit\"{a}t zu K\"{o}ln, Z\"{u}lpicher Str. 77, 50937 K\"{o}ln, Germany}
\author{T.~Burnus}
  \affiliation{II. Physikalisches Institut, Universit\"{a}t zu K\"{o}ln, Z\"{u}lpicher Str. 77, 50937 K\"{o}ln, Germany}
\author{N.~Hollmann}
  \affiliation{II. Physikalisches Institut, Universit\"{a}t zu K\"{o}ln, Z\"{u}lpicher Str. 77, 50937 K\"{o}ln, Germany}
\author{M.~Benomar}
  \affiliation{II. Physikalisches Institut, Universit\"{a}t zu K\"{o}ln, Z\"{u}lpicher Str. 77, 50937 K\"{o}ln, Germany}
\author{T.~Lorenz}
  \affiliation{II. Physikalisches Institut, Universit\"{a}t zu K\"{o}ln, Z\"{u}lpicher Str. 77, 50937 K\"{o}ln, Germany}
\author{A.~Tanaka}
  \affiliation{Department of Quantum Matter, ADSM, Hiroshima University, Higashi-Hiroshima 739-8530, Japan}
\author{H.-J.~Lin}
  \affiliation{National Synchrotron Radiation Research Center, 101 Hsin-Ann Road, Hsinchu 30077, Taiwan}
\author{H.~H.~Hsieh}
  \affiliation{Chung Cheng Institute of Technology, National Defense University, Taoyuan 335, Taiwan}
\author{C.~T.~Chen}
  \affiliation{National Synchrotron Radiation Research Center, 101 Hsin-Ann Road, Hsinchu 30077, Taiwan}
\author{L.~H.~Tjeng}
  \affiliation{II. Physikalisches Institut, Universit\"{a}t zu K\"{o}ln, Z\"{u}lpicher Str. 77, 50937 K\"{o}ln, Germany}

\date{\today}

\begin{abstract}
Using Co-$L_{2,3}$ and O-$K$ x-ray absorption spectroscopy, we
reveal that the charge ordering in La$_{1.5}$Sr$_{0.5}$CoO$_{4}$
involves high spin ($S$=3/2) Co$^{2+}$ and low spin ($S$=0)
Co$^{3+}$ ions. This provides evidence for the spin blockade
phenomenon as a source for the extremely insulating nature of the
La$_{2-x}$Sr$_{x}$CoO$_{4}$ series. The associated $e_g^2$ and
$e_g^0$ orbital occupation accounts for the large contrast in the
Co-O bond lengths, and in turn, the high charge ordering
temperature. Yet, the low magnetic ordering temperature is
naturally explained by the presence of the non-magnetic ($S$=0)
Co$^{3+}$ ions. From the identification of the bands we infer that
La$_{1.5}$Sr$_{0.5}$CoO$_{4}$ is a narrow band material.
\end{abstract}


\maketitle

Considerable research effort has been put in cobaltate materials
during the last decade in search for new phenomena and
extraordinary properties. A key aspect of cobaltates that
distinguish them from e.g. the manganates and cuprates
\cite{Imada98}, is the spin state degree of freedom of the
Co$^{3+/III}$ ions: it can be low spin (LS, $S$=0), high spin (HS,
$S$=2) and even intermediate spin (IS, $S$=1)
\cite{Sugano70,Goodenough71}. This aspect comes on top of the
orbital, spin (up/down) and charge degrees of freedom that already
make the manganates and cuprates so exciting. Indeed, numerous
cobaltate systems have been discovered with properties that
include giant magneto resistance \cite{Briceno95,Martin97},
superconductivity \cite{Takada03} and
ferro-ferri-antiferro-magnetic transitions with various forms of
charge, orbital and spin ordering
\cite{Fjellvag96,Vogt00,Loureiro00,Niitaka01,Burley03,Taskin03,
Kudasov06,Luetkens2008:PRL-101-017601}.
A new and exciting aspect in here is the recognition that the
so-called \emph{spin blockade} mechanism could be present and
responsible for several of those unusual properties
\cite{Maignan2004:PRL-93-026401}. If true, this would open up new
research opportunities since one could envision exploiting
explicitly this mechanism in materials design.

Here we focus on the La$_{2-x}$Sr$_x$CoO$_4$ system, which shows
quite peculiar transport and magnetic properties
\cite{Srivastava1963,Matsuura1988,Moritomo:PRB-55-R14725,Iguchi1998,
Itoh1999,Zaliznyak:PRL-85-4353,Zaliznyak:PRB-64-195117,Nils:NJP-10-023018,
Cwik2008:cond-mat,Sakiyama2008:PhysRevB-78-180406}. This material
is extremely insulating for a very wide range of $x$ values with
anomalously high activation energies for conductivity, very much
unlike the Mn, Ni, or Cu compounds
\cite{Imada98,Moritomo:PRB-55-R14725,Munnings:SSI-177-1849}. The
commensurate antiferromagnetic (AF) state remains stable up to a
surprisingly high value of $x$=0.3
\cite{Cwik2008:cond-mat,Sakiyama2008:PhysRevB-78-180406}. Charge
ordering (CO) and spin ordering (SO) at half doping involve quite
different transition temperatures, namely $T_{CO}\sim750$ K and
$T_{SO}\leq30$ K, respectively. This constitutes a ratio of 25,
which is extraordinary since it is an order of magnitude larger
than in the Mn and Ni materials
\cite{Imada98,Zaliznyak:PRL-85-4353,Kajimoto2003:PRB-67-014511}.

It was already reported that the SO in the
La$_{1.5}$Sr$_{0.5}$CoO$_4$ composition involves non-magnetic
Co$^{3+}$ ions with the claim that these Co$^{3+}$ ions are in the
IS state and become non-magnetic due to strong planar anisotropy
driven quenching of the spin angular momentum below the $T_{SO}$
\cite{Zaliznyak:PRL-85-4353,Zaliznyak:PRB-64-195117}. Here we go
one step further. Using soft x-ray absorption spectroscopy (XAS)
we are able to show unambiguously that the Co$^{3+}$ ions are in
the LS ($S$=0) state, both below and above $T_{SO}$. Together with
the verification that the Co$^{2+}$ ions are HS ($S$=3/2), we
establish that the spin blockade mechanism is active. The highly
insulating character of the material over a wide range of
temperatures is thus explained. Important is that the associated
$e_g^0$ and $e_g^2$ orbital occupation ordering leads to a large
difference in the Co-O bond lengths and the high CO temperature.
At the same time, the low SO temperature follows naturally from
the presence of the truely non-magnetic ($S$=0) Co$^{3+}$ ions.

Single crystals of La$_{1.5}$Sr$_{0.5}$CoO$_{4}$ have been grown
by the travelling floating-zone method and characterized using
magnetic and resistivity measurements \cite{Nils:NJP-10-023018}.
The polarization dependent XAS experiments were performed at the
Dragon beamline of the National Synchrotron Radiation Research
Center (NSRRC) in Taiwan. The Co-$L_{2,3}$ spectra were taken in
the total electron yield (TEY) mode and the O-$K$ also in the
fluorescence yield (FY) mode with a photon energy resolution of
0.3 and 0.2 eV, respectively. The degree of linear polarization of
the incident light was ~99\%. The crystals were mounted with the
${\mathbf c}$-axis perpendicular to the Poynting vector of the
light. By rotating the sample around this Poynting vector, the
polarization of the electric field can be varied continuously from
$\mathbf{E}\parallel\mathbf{c}$ to $\mathbf{E}\perp\mathbf{c}$.
The crystals were cleaved parallel to the ${\mathbf c}$-axis in a
vacuum of $10^{-10}$ mbar.

\begin{figure}
    \centering
    \includegraphics[width=6cm]{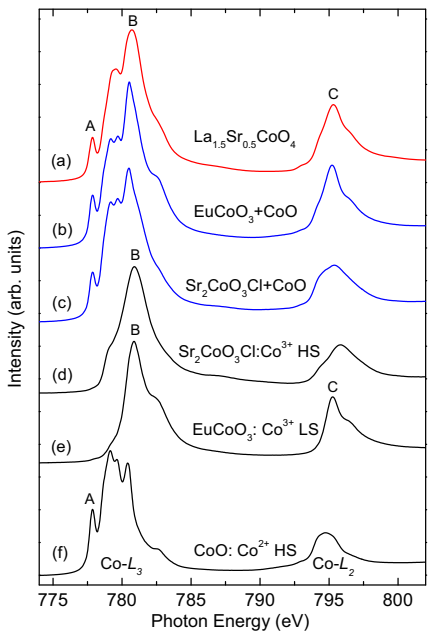}
    \caption{(color online) Isotropic Co-$L_{2,3}$ XAS spectra of
    (a) La$_{1.5}$Sr$_{0.5}$CoO$_{4}$, (b) the HS-CoO plus LS-EuCoO$_{3}$ scenario,
    (c) the HS-CoO plus HS-Sr$_{2}$CoO$_{3}$Cl scenario, (d) Sr$_{2}$CoO$_{3}$Cl,
    (e) EuCoO$_{3}$, and (f) CoO, all taken at 300 K.}
    \label{Fig-ISO-XAS}
\end{figure}

Fig.~\ref{Fig-ISO-XAS} depicts the room temperature isotropic
Co-$L_{2,3}$ XAS spectra of La$_{1.5}$Sr$_{0.5}$CoO$_{4}$ together
with those of Sr$_{2}$CoO$_{3}$Cl, EuCoO$_{3}$, and CoO serving as
the HS-Co$^{3+}$, the LS-Co$^{3+}$, and the HS-Co$^{2+}$
references, respectively \cite{hu2004:PRL-92-207402}. The spectra
are dominated by the Co $2p$ core-hole spin-orbit coupling which
splits the spectrum roughly in two parts, namely the $L_{3}$
($h\nu \approx$ 776-784 eV) and $L_{2}$ ($h\nu \approx$ 793-797
eV) white lines regions. The line shape strongly depends on the
multiplet structure given by the Co 3$d$-3$d$ and 2$p$-3$d$
Coulomb and exchange interactions, as well as by the local crystal
fields and the hybridization with the O 2$p$ ligands. Unique to
soft x-ray absorption is that the dipole selection rules are very
sensitive in determining which of the 2$p^{5}$3$d^{n+1}$ final
states can be reached and with what intensity, starting from a
particular 2$p^{6}$3$d^{n}$ initial state ($n= 6$ for Co$^{3+}$
and $n =7$ for Co$^{2+}$) \cite{DeGroot1994,Tanaka1994}. This
makes the technique extremely sensitive to the symmetry of the
initial state, i.e., the spin, orbital and valence states of the
ions \cite{Abbate1993:PRB-47-16124,hu2004:PRL-92-207402,
haverkort2006:PRL-97-176405,burnus2006:PRB-74-245111,
csiszar2005:PRL-950187205,Mitra2003:PRB.67.092404}.

The spectra of the reference samples show quite different
multiplet structures with peaks at quite different energies. In
particular, the lowest energy peak (A) of the CoO at 777 eV can be
taken as a characteristic for the presence of an octahedral
Co$^{2+}$ ion since it has an energy well below that of any
Co$^{3+}$ or Co$^{4+}$ (not shown here) features. Also very
characteristic is the dominant peak (B) at 781 eV of a Co$^{3+}$
ion. In comparing the spectrum of EuCoO$_{3}$ with that of
Sr$_{2}$CoO$_{3}$Cl, one can, for example, notice that the LS
Co$^{3+}$ ion has a higher intensity at the $L_2$ edge with a
rather sharp peak (C) at 795 eV. Focussing now on the
La$_{1.5}$Sr$_{0.5}$CoO$_{4}$ spectrum, one can clearly observe,
among others, a low energy peak (A) at 777 eV, a dominant peak (B)
at 781 eV, and a sharply peaked $L_2$ edge (C) at 795 eV. This
strongly hints towards the presence of not only Co$^{2+}$ and
Co$^{3+}$ ions in this material, but especially that the Co$^{3+}$
is LS. To verify this, we have carried out a simple simulation by
making a superposition of the as-measured CoO and EuCoO$_3$
spectra and compare the result with the spectrum of
La$_{1.5}$Sr$_{0.5}$CoO$_{4}$. We see from Fig.~\ref{Fig-ISO-XAS}
that the simulation is almost perfect. As a counter check, we also
made a superposition of the CoO and Sr$_{2}$CoO$_{3}$Cl spectra,
and observe noticeable discrepancies between this superposition
and the spectrum of La$_{1.5}$Sr$_{0.5}$CoO$_{4}$, especially in
the $L_2$ edge region.

\begin{figure}
    \centering
    \includegraphics[width=6.5cm]{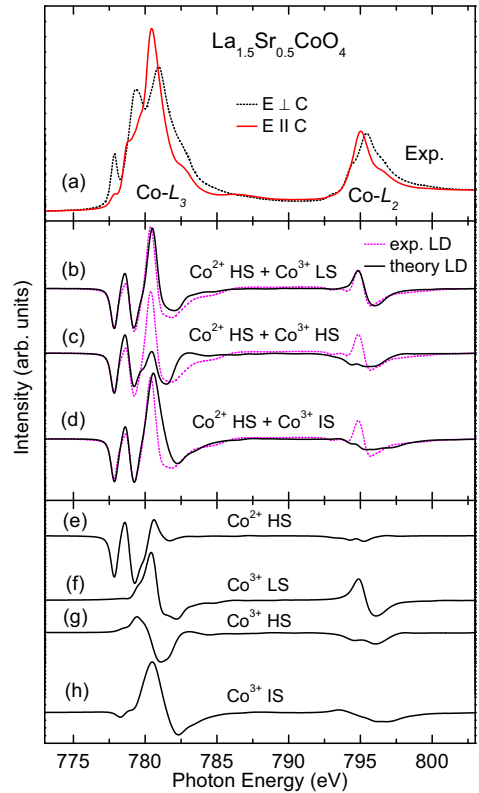}
    \caption{(color online) (a) Experimental polarization-dependent
    Co-$L_{2,3}$ XAS of La$_{1.5}$Sr$_{0.5}$CoO$_{4}$ for
    $\mathbf{E}\parallel\mathbf{c}$ (red line) and
    $\mathbf{E}\perp\mathbf{c}$ (black dotted line); (b), (c), and
    (d) the experimental linear dichroism (magenta dotted lines) together with
    the theoretical one (black lines) calculated for
    scenarios with HS-Co$^{2+}$ and LS-, HS-, or IS-Co$^{3+}$, respectively.
    The calculated linear dichroism for the HS-Co$^{2+}$, LS-, HS-, and
    IS-Co$^{3+}$ is also shown individually in (e), (f), (g), and (h).}
    \label{Fig-Cal-XAS}
\end{figure}

In order to further confirm that the Co$^{3+}$ ions are in the LS
state, we have measured the polarization dependence of the
Co-$L_{2,3}$ XAS of La$_{1.5}$Sr$_{0.5}$CoO$_{4}$ and have
simulated the spectra using the successful configuration
interaction cluster model that includes the full atomic multiplet
theory and the hybridization with the O $2p$ ligands
\cite{DeGroot1994,Tanaka1994}. For this we use parameter values
typical for Co$^{2+}$ and Co$^{3+}$ systems
\cite{hu2004:PRL-92-207402,haverkort2006:PRL-97-176405,
burnus2006:PRB-74-245111,CIparameters}. The Co $3d$ to O $2p$
transfer integrals were adapted for the various Co-O bond lengths
\cite{Cwik2007} according to Harrison's prescription
\cite{Harrison1989}. The crystal field parameters were determined
from constrained density-functional calculations using the
linearized augmented plane wave plus local orbital method
\cite{WIEN2k} and the electron-correlation corrected
local-density-approximation (LDA+\emph{U} with $U$=5 eV)
\cite{Anisimov1993:PhysRevB-48-16929}. The cluster model
calculations were done using XTLS 8.3 \cite{Tanaka1994}.

The top panel of Fig.~\ref{Fig-Cal-XAS} (a) shows the experimental
Co-$L_{2,3}$ XAS of La$_{1.5}$Sr$_{0.5}$CoO$_{4}$ taken with
$\mathbf{E}\parallel\mathbf{c}$ (red) and
$\mathbf{E}\perp\mathbf{c}$ (black). The experimental linear
dichroic (LD) signal, defined as the difference between two
polarizations, is displayed in the middle panel (magenta),
together with the simulated LD spectra (black) for the Co$^{3+}$
ion in the (b) LS, (c) HS, and (d) IS state, respectively, while
the Co$^{2+}$ ion is kept in the HS. One can observe from the
middle panel that the LS-Co$^{3+}$ scenario nicely reproduces all
features of the experimental LD spectrum. In contrast, the
HS-Co$^{3+}$ assumption would lead to significant discrepancies,
not only at the $L_3$ but also at the $L_2$ edge. Also the
IS-Co$^{3+}$ scenario gives less than satisfactory fit to the
experimental spectrum: the $L_2$ edge structure cannot be
reproduced at all. More insight can be gained from the lower panel
of Fig.~\ref{Fig-Cal-XAS} in which the calculated LD spectra are
individually depicted. The Co$^{2+}$ LD spectrum (e) shows only a
modest modulation at the L$_2$ edge, and so does the HS-Co$^{3+}$
(g) and the IS-Co$^{3+}$ (h) as well. Only the LS-Co$^{3+}$ (f)
displays the strong LD modulation as experimentally observed for
the $L_2$ region.

So far the spin-state issue in the La$_{2-x}$Sr$_x$CoO$_4$ system
has been addressed using magnetic susceptibility
\cite{Srivastava1963,Matsuura1988,Moritomo:PRB-55-R14725,Nils:NJP-10-023018}
or NMR measurements \cite{Itoh1999}, with conflicting results: all
three possible scenarios (HS, IS, LS) have been proposed for the
Co$^{3+}$ ion. Interestingly, neutron scattering experiments have
revealed the presence of non-magnetic Co ions in the SO state of
La$_{1.5}$Sr$_{0.5}$CoO$_4$ \cite{Zaliznyak:PRL-85-4353}. Yet it
was claimed that these Co$^{3+}$ ions are in the IS state and that
they become non-magnetic as a result of the quenching of the spin
angular momentum due to the strong planar anisotropy in the SO
state \cite{Zaliznyak:PRL-85-4353,Zaliznyak:PRB-64-195117}.

\begin{figure}
    \centering
    \includegraphics[width=6.5cm]{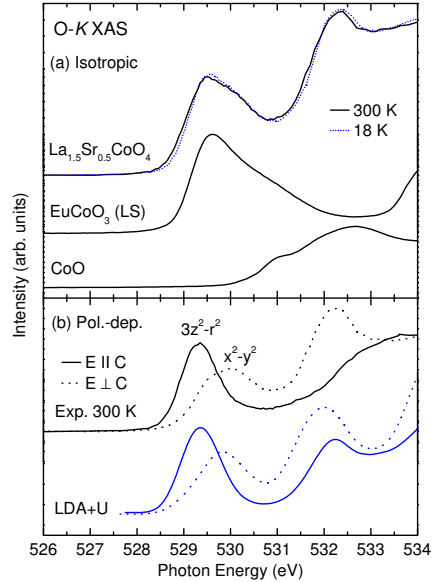}
    \caption{(color online) (a) Isotropic O-$K$ XAS of
    La$_{1.5}$Sr$_{0.5}$CoO$_{4}$ taken at 18 K (blue dotted line) and
    300 K (black solid line), together with those of EuCoO3 (LS) and CoO.
    (b) Polarization-dependent O-$K$ XAS, taken with
    $\mathbf{E}$$\parallel$$\mathbf{c}$ (solid line) and
    $\mathbf{E}$$\perp$$\mathbf{c}$ (dotted line),
    together with the LDA+\emph{U} unoccupied O $2p$ density
    of states (blue curves).}
    \label{Fig-O-1s}
\end{figure}

Our XAS data yield new and crucial information. The Co$^{3+}$ ions
are indeed non-magnetic, but because of a completely different
reason: they are in the LS ($S$=0) state. The XAS data were taken
at room temperature, i.e. above $T_{SO}$. We expect that this LS
($S$=0) state is also stable below $T_{SO}$. We have verified this
using yet another spectroscopic method, namely the O-$K$ XAS. The
top panel of Fig.~\ref{Fig-O-1s} (a) displays the isotropic
spectrum of La$_{1.5}$Sr$_{0.5}$CoO$_4$ taken at 300 K and 18 K.
These spectra were taken in the FY method to avoid charging
problems which otherwise could occur at low sample temperatures
when using the TEY. One can clearly see that the lineshape of the
529.5 eV pre-edge feature at both temperatures is very similar to
that of EuCoO$_3$, a reliable LS-Co$^{3+}$ reference
\cite{hu2004:PRL-92-207402}. The temperature dependence, if any,
is only a slight broadening with increasing temperature, and
certainly not the massive change of the spectral lineshape (over a
range of $\approx$1 eV) as reported for LaCoO$_3$ in going from
the low-temperature non-magnetic to the high-temperature
paramagnetic state \cite{Abbate1993:PRB-47-16124}.

The presence of LS-Co$^{3+}$ ions provides a natural explanation
for the rapid lowering of the $T_{SO}$ when doping La$_2$CoO$_4$
with Sr. The number of paths with strong superexchange
interactions between the HS-Co$^{2+}$ ions are simply reduced when
non-magnetic Co ions are introduced. It should therefore be of no
surprise that the $T_{SO}$ drops from 275 K for La$_2$CoO$_4$ to a
low 30 K for La$_{1.5}$Sr$_{0.5}$CoO$_4$ in which the
Co$^{2+}$/Co$^{3+}$ ions are checkerboard ordered.

The presence of LS-Co$^{3+}$ ions in between HS-Co$^{2+}$ is also
exciting since it gives a beautiful explanation for the highly
insulating behavior in La$_{2-x}$Sr$_{x}$CoO$_4$ despite the heavy
doping. If one considers a pair of neighboring HS-Co$^{2+}$
($S$=3/2) and LS-Co$^{3+}$ ($S$=0) ions, then one can directly see
that the hopping of a charge carrier will not result in an
identical pair with just the charges interchanged. Instead, the
pair created will have a completely different spin state
configuration: the hopping of a spin 1/2 particle will produce a
IS-Co$^{3+}$ ($S$=1) and LS-Co$^{2+}$ ($S$=1/2) final state.
Although such a hopping does not involve the Coulomb energy $U$,
it does cost a significant amount of energy since the energy
difference between a HS- and LS-Co$^{2+}$ as well as between a LS-
and IS-Co$^{3+}$ could amount to several tenths of an eV
\cite{Sugano70}. The hopping of charge carriers between these two
types of Co ions is thus severely suppressed. This type of
suppression of the conductivity, known as spin-blockade, has been
proposed to explain the thermoelectric power of
HoBaCo$_2$O$_{5.5}$ \cite{Maignan2004:PRL-93-026401}. We note that
the spin-blockade mechanism is effective in
La$_{1.5}$Sr$_{0.5}$CoO$_{4}$ both below and above the $T_{SO}$
since the LS state of the Co$^{3+}$ is independent of the SO. This
explains why the activation energies for the conductivity remain
high even at elevated temperatures.

The spin state contrast which goes along with the
Co$^{2+}$/Co$^{3+}$ charge state, has also consequences for the
stability of the CO. The orbital occupation of the HS-Co$^{2+}$ is
predominantly $t_{2g}^{5}$$e_{g}^{2}$ while that of the
LS-Co$^{3+}$ is mainly $t_{2g}^{6}$$e_{g}^{0}$, as we have
verified from our cluster calculations. There is thus also an
extremely strong $e_g$ orbital occupation contrast, even when
taking into account that the LS-Co$^{3+}$ is more covalent than
the HS-Co$^{2+}$ \cite{Saitoh:PhysRevB-55-4257}. With the $e_g$
orbitals pointing towards the oxygens and the $t_{2g}$ in between,
this leads to significant differences in the local Co-O distances
for the two different ions, as indeed revealed by diffraction
experiments
\cite{Zaliznyak:PRL-85-4353,Cwik2008:cond-mat,Sakiyama2008:PhysRevB-78-180406}.
This in turn causes strong localization and stabilization of the
charge contrast, thereby explaining the very high $T_{CO}\sim750$
K in La$_{1.5}$Sr$_{0.5}$CoO$_{4}$.

There is yet another aspect which makes this system to be highly
insulating. For this we look in more detail to the energetics of
the unoccupied states of the Co$^{3+}$. The bottom panel of
Fig.~\ref{Fig-O-1s} (b) shows the polarization dependence of the
O-$K$ XAS. Here we are mainly interested in the lowest lying
states, i.e. the unoccupied states of the Co$^{3+}$ which can be
reached when an electron is transferred from a neighboring
Co$^{2+}$. These states are located in the so-called pre-edge
region between 528-531 eV \cite{Abbate1993:PRB-47-16124}. Using
$\mathbf{E}\parallel\mathbf{c}$ light (solid line) we can reach
the unoccupied O-$2p_z$--Co-$3d_{3z^2-r^2}$ hybridized band, and
found it to be peaked at 529.3 eV. With
$\mathbf{E}\perp\mathbf{c}$ light (dotted line), we see that the
peak of the mixed O-$2p_{x,y}$--Co-$3d_{x^2-y^2}$ band lies 0.7 eV
higher, at 530.0 eV. The lowest energy states at 528.5 eV are of
almost pure $3z^{2}$--$r^{2}$ nature. These assignments are very
well supported by the LDA+\emph{U} calculations: not only the
larger intensity of the $3z^{2}$--$r^{2}$ (solid line) in
comparison to that of the $x^2$--$y^2$ state (dotted line) is
reproduced but also the relative energy positions.

This is an important finding since it establishes that the band
available for conduction, namely the one with the lowest energy,
is of $3z^{2}$$-$$r^{2}$ origin. This is very different from the
cuprate, nickelate and manganate systems, where the relevant band
is formed out of the $x^2$$-$$y^2$ orbital. The consequences are
obvious: with the $3z^{2}$$-$$r^{2}$ orbital having much smaller
overlap with the in-plane O-$2p_{x,y}$ ligands than the
$x^2$$-$$y^2$, the conduction band gets much narrower, with the
result that the spin-blockade effect together with the strong
coupling of the Co-O distances with the charge/spin state of the
Co ions become the dominant interactions to make
La$_{2-x}$Sr$_{x}$CoO$_4$ highly insulating over a wide range of
$x$.

We note that the observed and calculated 0.7 eV energy shift
between the $3z^2$$-$$r^2$ (solid line) and the $x^2$$-$$y^2$
(dotted line) states in Fig.~\ref{Fig-O-1s}(b) reflects the large
tetragonal distortion of about 10\% \cite{Cwik2007}. This large
crystal field splitting in the $e_g$ orbital is in fact the cause
for the strong modulation in the LD spectrum at the $L_2$ edge for
the LS-Co$^{3+}$ ion as depicted in the bottom panel of
Fig.~\ref{Fig-Cal-XAS}, curve (f); without such an energy
splitting which is felt by the electron promoted from the Co $2p$
core, one would not observe any dichroism in the otherwise highly
symmetric, closed-shell $t_{2g}^{6}$ ground state.

To summarize, we find using soft x-ray absorption spectroscopy at
the Co-$L_{2,3}$ and O-$K$ edges that the charge ordering in
La$_{1.5}$Sr$_{0.5}$CoO$_{4}$ involves high spin ($S$=3/2)
Co$^{2+}$ and low spin ($S$=0) Co$^{3+}$ ions. We infer that the
spin blockade mechanism is active here and that there is a strong
coupling between the local Co-O distances and the charge/spin
state of the ions. The crystal field scheme for the Co$^{3+}$ ion
caused by the tetragonal distortion makes the conduction band
extremely narrow. All these factors provide an explanation for the
highly insulating properties as well as for the very low $T_{SO}$
and the exceptionally high $T_{CO}$.

We thank Lucie Hamdan and Matthias Cwik for their skillful
technical assistance. The research in Cologne is supported by the
Deutsche Forschungsgemeinschaft through SFB 608.


\begin{thebibliography}{99}

\expandafter\ifx\csname
natexlab\endcsname\relax\def\natexlab#1{#1}\fi
\expandafter\ifx\csname bibnamefont\endcsname\relax
  \def\bibnamefont#1{#1}\fi
\expandafter\ifx\csname bibfnamefont\endcsname\relax
  \def\bibfnamefont#1{#1}\fi
\expandafter\ifx\csname citenamefont\endcsname\relax
  \def\citenamefont#1{#1}\fi
\expandafter\ifx\csname url\endcsname\relax
  \def\url#1{\texttt{#1}}\fi
\expandafter\ifx\csname
urlprefix\endcsname\relax\def\urlprefix{URL }\fi
\providecommand{\bibinfo}[2]{#2}
\providecommand{\eprint}[2][]{\url{#2}}


\bibitem[{\citenamefont{Imada et~al.}(1998)\citenamefont{Imada, Fujimori, and
  Tokura}}]{Imada98}
\bibinfo{author}{\bibfnamefont{M.}~\bibnamefont{Imada}},
  \bibinfo{author}{\bibfnamefont{A.}~\bibnamefont{Fujimori}}, \bibnamefont{and}
  \bibinfo{author}{\bibfnamefont{Y.}~\bibnamefont{Tokura}},
  \bibinfo{journal}{Rev. Mod. Phys.} \textbf{\bibinfo{volume}{70}},
  \bibinfo{pages}{1039} (\bibinfo{year}{1998}).


\bibitem[{\citenamefont{Sugano et~al.}(1970)\citenamefont{Sugano, Tanabe, and
  Kamimura}}]{Sugano70}
\bibinfo{author}{\bibfnamefont{S.}~\bibnamefont{Sugano}},
  \bibinfo{author}{\bibfnamefont{Y.}~\bibnamefont{Tanabe}}, \bibnamefont{and}
  \bibinfo{author}{\bibfnamefont{H.}~\bibnamefont{Kamimura}},
  \emph{\bibinfo{title}{Multiplets of Transition-Metal Ions in Crystals}}
  (\bibinfo{publisher}{Academic}, \bibinfo{address}{New York},
  \bibinfo{year}{1970}).


\bibitem[{\citenamefont{Goodenough}(1971)}]{Goodenough71}
\bibinfo{author}{\bibfnamefont{J.~B.} \bibnamefont{Goodenough}}, in
  \emph{\bibinfo{booktitle}{Progress in Solid State Chemistry}}, edited by
  \bibinfo{editor}{\bibfnamefont{H.}~\bibnamefont{Reiss}}
  (\bibinfo{publisher}{Pergamon}, \bibinfo{address}{Oxford},
  \bibinfo{year}{1971}), vol.~\bibinfo{volume}{5}.


\bibitem[{\citenamefont{Brice\~{n}o et~al.}(1995)}]{Briceno95}
\bibinfo{author}{\bibfnamefont{G.}~\bibnamefont{Brice\~{n}o}}
  \bibnamefont{\emph{et~al.}}, \bibinfo{journal}{Science}
  \textbf{\bibinfo{volume}{270}}, \bibinfo{pages}{273} (\bibinfo{year}{1995}).


\bibitem[{\citenamefont{Martin et~al.}(1997)}]{Martin97}
\bibinfo{author}{\bibfnamefont{C.}~\bibnamefont{Martin}}
  \bibnamefont{\emph{et~al.}}, \bibinfo{journal}{Appl. Phys. Lett.}
  \textbf{\bibinfo{volume}{71}}, \bibinfo{pages}{1421} (\bibinfo{year}{1997}).


\bibitem[{\citenamefont{Takada et~al.}(2003)}]{Takada03}
\bibinfo{author}{\bibfnamefont{K.}~\bibnamefont{Takada}}
  \bibnamefont{\emph{et~al.}}, \bibinfo{journal}{Nature (London)}
  \textbf{\bibinfo{volume}{422}}, \bibinfo{pages}{53} (\bibinfo{year}{2003}).


\bibitem[{\citenamefont{Fjellv{\aa}g et~al.}(1996)}]{Fjellvag96}
\bibinfo{author}{\bibfnamefont{H.}~\bibnamefont{Fjellv{\aa}g}}
  \bibnamefont{\emph{et~al.}}, \bibinfo{journal}{J. Solid State Chem.}
  \textbf{\bibinfo{volume}{124}}, \bibinfo{pages}{190} (\bibinfo{year}{1996}).

\bibitem[{\citenamefont{Vogt et~al.}(2000)}]{Vogt00}
\bibinfo{author}{\bibfnamefont{T.}~\bibnamefont{Vogt}}
  \bibnamefont{\emph{et~al.}}, \bibinfo{journal}{Phys. Rev. Lett.}
  \textbf{\bibinfo{volume}{84}}, \bibinfo{pages}{2969} (\bibinfo{year}{2000}).


\bibitem[{\citenamefont{Loureiro et~al.}(2000)\citenamefont{Loureiro, Felser,
  Huang, and Cava}}]{Loureiro00}
\bibinfo{author}{\bibfnamefont{S.}~\bibnamefont{Loureiro}},
  \bibinfo{author}{\bibfnamefont{C.}~\bibnamefont{Felser}},
  \bibinfo{author}{\bibfnamefont{Q.}~\bibnamefont{Huang}}, \bibnamefont{and}
  \bibinfo{author}{\bibfnamefont{R.}~\bibnamefont{Cava}},
  \bibinfo{journal}{Chem. Mater.} \textbf{\bibinfo{volume}{12}},
  \bibinfo{pages}{3181} (\bibinfo{year}{2000}).


\bibitem[{\citenamefont{Niitaka et~al.}(2001)}]{Niitaka01}
\bibinfo{author}{\bibfnamefont{S.}~\bibnamefont{Niitaka}}
  \bibnamefont{\emph{et~al.}}, \bibinfo{journal}{Phys. Rev. Lett.}
  \textbf{\bibinfo{volume}{87}}, \bibinfo{pages}{177202}
  (\bibinfo{year}{2001}).


\bibitem[{\citenamefont{Burley et~al.}(2003)}]{Burley03}
\bibinfo{author}{\bibfnamefont{J.~C.} \bibnamefont{Burley}}
  \bibnamefont{\emph{et~al.}}, \bibinfo{journal}{J. Solid State Chem.}
  \textbf{\bibinfo{volume}{170}}, \bibinfo{pages}{339} (\bibinfo{year}{2003}).

\bibitem[{\citenamefont{Taskin et~al.}(2003)\citenamefont{Taskin, Lavrov, and
  Ando}}]{Taskin03}
\bibinfo{author}{\bibfnamefont{A.~A.} \bibnamefont{Taskin}},
  \bibinfo{author}{\bibfnamefont{A.~N.} \bibnamefont{Lavrov}},
  \bibnamefont{and} \bibinfo{author}{\bibfnamefont{Y.}~\bibnamefont{Ando}},
  \bibinfo{journal}{Phys. Rev. Lett.} \textbf{\bibinfo{volume}{90}},
  \bibinfo{pages}{227201} (\bibinfo{year}{2003}).

\bibitem[{\citenamefont{Kudasov}(2006)}]{Kudasov06}
\bibinfo{author}{\bibfnamefont{Y.~B.} \bibnamefont{Kudasov}},
  \bibinfo{journal}{Phys. Rev. Lett.} \textbf{\bibinfo{volume}{96}},
  \bibinfo{eid}{027212} (\bibinfo{year}{2006}).

\bibitem[{\citenamefont{Luetkens et~al.}(2008)}]{Luetkens2008:PRL-101-017601}
\bibinfo{author}{\bibfnamefont{H.} \bibnamefont{Luetkens}}
  \bibnamefont{\emph{et~al.}}, \bibinfo{journal}{Phys. Rev. Lett.}
  \textbf{\bibinfo{volume}{101}}, \bibinfo{eid}{017601} (\bibinfo{year}{2008}).

\bibitem[{\citenamefont{Maignan et~al.}(2004)}]{Maignan2004:PRL-93-026401}
\bibinfo{author}{\bibfnamefont{A.}~\bibnamefont{Maignan}}
  \bibnamefont{\emph{et~al.}}, \bibinfo{journal}{Phys. Rev. Lett.}
  \textbf{\bibinfo{volume}{93}}, \bibinfo{pages}{026401}
  (\bibinfo{year}{2004}).


\bibitem[{\citenamefont{Srivastava}(1963)}]{Srivastava1963}
\bibinfo{author}{\bibfnamefont{K.~G.} \bibnamefont{Srivastava}},
  \bibinfo{journal}{Phys. Lett.} \textbf{\bibinfo{volume}{4}},
  \bibinfo{pages}{55} (\bibinfo{year}{1963}).


\bibitem[{\citenamefont{Matsuura et~al.}(1988)}]{Matsuura1988}
\bibinfo{author}{\bibfnamefont{T.}~\bibnamefont{Matsuura}}
  \bibnamefont{\emph{et~al.}}, \bibinfo{journal}{J. Phys. Chem. Solids}
  \textbf{\bibinfo{volume}{49}}, \bibinfo{pages}{1403}
  (\bibinfo{year}{1988}); \textit{ibid.} \textbf{\bibinfo{volume}{49}},
  \bibinfo{pages}{1409} (\bibinfo{year}{1988}).


\bibitem[{\citenamefont{Moritomo et~al.}(1997)\citenamefont{Moritomo, Higashi,
  Matsuda, and Nakamura}}]{Moritomo:PRB-55-R14725}
\bibinfo{author}{\bibfnamefont{Y.}~\bibnamefont{Moritomo}},
  \bibinfo{author}{\bibfnamefont{K.}~\bibnamefont{Higashi}},
  \bibinfo{author}{\bibfnamefont{K.}~\bibnamefont{Matsuda}}, \bibnamefont{and}
  \bibinfo{author}{\bibfnamefont{A.}~\bibnamefont{Nakamura}},
  \bibinfo{journal}{Phys. Rev. B} \textbf{\bibinfo{volume}{55}},
  \bibinfo{pages}{R14725} (\bibinfo{year}{1997}).


\bibitem[{\citenamefont{Iguchi et~al.}(1998)\citenamefont{Iguchi, Nakatsugawa,
  and Futakuchi}}]{Iguchi1998}
\bibinfo{author}{\bibfnamefont{E.}~\bibnamefont{Iguchi}},
  \bibinfo{author}{\bibfnamefont{H.}~\bibnamefont{Nakatsugawa}},
  \bibnamefont{and}
  \bibinfo{author}{\bibfnamefont{K.}~\bibnamefont{Futakuchi}},
  \bibinfo{journal}{J. Solid State Chem.} \textbf{\bibinfo{volume}{139}},
  \bibinfo{pages}{176} (\bibinfo{year}{1998}).


\bibitem[{\citenamefont{Itoh et~al.}(1999)\citenamefont{Itoh, Mori, Moritomo,
  and Nakamura}}]{Itoh1999}
\bibinfo{author}{\bibfnamefont{M.}~\bibnamefont{Itoh}},
  \bibinfo{author}{\bibfnamefont{M.}~\bibnamefont{Mori}},
  \bibinfo{author}{\bibfnamefont{Y.}~\bibnamefont{Moritomo}}, \bibnamefont{and}
  \bibinfo{author}{\bibfnamefont{A.}~\bibnamefont{Nakamura}},
  \bibinfo{journal}{Physica B} \textbf{\bibinfo{volume}{259-261}},
  \bibinfo{pages}{997} (\bibinfo{year}{1999}).


\bibitem[{\citenamefont{Zaliznyak et~al.}(2000)}]{Zaliznyak:PRL-85-4353}
\bibinfo{author}{\bibfnamefont{I.~A.} \bibnamefont{Zaliznyak}}
  \bibnamefont{\emph{et~al.}}, \bibinfo{journal}{Phys. Rev. Lett.}
  \textbf{\bibinfo{volume}{85}}, \bibinfo{pages}{4353} (\bibinfo{year}{2000}).


\bibitem[{\citenamefont{Zaliznyak et~al.}(2001)\citenamefont{Zaliznyak,
  Tranquada, Erwin, and Moritomo}}]{Zaliznyak:PRB-64-195117}
\bibinfo{author}{\bibfnamefont{I.~A.} \bibnamefont{Zaliznyak}},
  \bibinfo{author}{\bibfnamefont{J.~M.} \bibnamefont{Tranquada}},
  \bibinfo{author}{\bibfnamefont{R.}~\bibnamefont{Erwin}}, \bibnamefont{and}
  \bibinfo{author}{\bibfnamefont{Y.}~\bibnamefont{Moritomo}},
  \bibinfo{journal}{Phys. Rev. B} \textbf{\bibinfo{volume}{64}},
  \bibinfo{pages}{195117} (\bibinfo{year}{2001}).


\bibitem[{\citenamefont{Hollmann et~al.}(2008)}]{Nils:NJP-10-023018}
\bibinfo{author}{\bibfnamefont{N.}~\bibnamefont{Hollmann}}
  \bibnamefont{\emph{et~al.}}, \bibinfo{journal}{New J. Phys.}
  \textbf{\bibinfo{volume}{10}}, \bibinfo{pages}{023018}
  (\bibinfo{year}{2008}).

\bibitem[{\citenamefont{Cwik et~al.}(2008)}]{Cwik2008:cond-mat}
\bibinfo{author}{\bibfnamefont{M.}~\bibnamefont{Cwik}}
  \bibnamefont{\emph{et~al.}}, \bibinfo{journal}{arXiv:0808.0106v1}
  (\bibinfo{year}{2008}).

\bibitem[{\citenamefont{Sakiyama et~al.}(1993)}]{Sakiyama2008:PhysRevB-78-180406}
\bibinfo{author}{\bibfnamefont{N.} \bibnamefont{Sakiyama}}
  \bibnamefont{\emph{et~al.}}, \bibinfo{journal}{Phys. Rev. B}
  \textbf{\bibinfo{volume}{78}}, \bibinfo{pages}{180406(R)} (\bibinfo{year}{2008}).

\bibitem[{\citenamefont{Munnings et~al.}(2006)}]{Munnings:SSI-177-1849}
\bibinfo{author}{\bibfnamefont{C.~N.} \bibnamefont{Munnings}}
  \bibnamefont{\emph{et~al.}}, \bibinfo{journal}{Solid State Ionics}
  \textbf{\bibinfo{volume}{177}}, \bibinfo{pages}{1849} (\bibinfo{year}{2006}).

\bibitem[{\citenamefont{Kajimoto et~al.}(2003)\citenamefont{Kajimoto, Ishizaka,
  Yoshizawa, and Tokura}}]{Kajimoto2003:PRB-67-014511}
\bibinfo{author}{\bibfnamefont{R.}~\bibnamefont{Kajimoto}},
  \bibinfo{author}{\bibfnamefont{K.}~\bibnamefont{Ishizaka}},
  \bibinfo{author}{\bibfnamefont{H.}~\bibnamefont{Yoshizawa}},
  \bibnamefont{and} \bibinfo{author}{\bibfnamefont{Y.}~\bibnamefont{Tokura}},
  \bibinfo{journal}{Phys. Rev. B} \textbf{\bibinfo{volume}{67}},
  \bibinfo{pages}{014511} (\bibinfo{year}{2003}).


\bibitem[{\citenamefont{Hu et~al.}(2004)}]{hu2004:PRL-92-207402}
\bibinfo{author}{\bibfnamefont{Z.}~\bibnamefont{Hu}}
  \bibnamefont{\emph{et~al.}}, \bibinfo{journal}{Phys. Rev. Lett.}
  \textbf{\bibinfo{volume}{92}}, \bibinfo{eid}{207402} (\bibinfo{year}{2004}).


\bibitem[{\citenamefont{de~Groot}(1994)}]{DeGroot1994}
\bibinfo{author}{\bibfnamefont{F.~M.~F.} \bibnamefont{de~Groot}},
  \bibinfo{journal}{J. Electron Spectrosc. Relate. Phenom.}
  \textbf{\bibinfo{volume}{67}}, \bibinfo{pages}{529} (\bibinfo{year}{1994}).

\bibitem[{\citenamefont{Tanaka and Jo}(1994)}]{Tanaka1994}
\bibinfo{author}{\bibfnamefont{A.}~\bibnamefont{Tanaka}} \bibnamefont{and}
  \bibinfo{author}{\bibfnamefont{T.}~\bibnamefont{Jo}}, \bibinfo{journal}{J.
  Phys. Soc. Jpn.} \textbf{\bibinfo{volume}{63}}, \bibinfo{pages}{2788}
  (\bibinfo{year}{1994}).


\bibitem[{\citenamefont{Abbate et~al.}(1993)}]{Abbate1993:PRB-47-16124}
\bibinfo{author}{\bibfnamefont{M.}~\bibnamefont{Abbate}}
  \bibnamefont{\emph{et~al.}}, \bibinfo{journal}{Phys. Rev. B}
  \textbf{\bibinfo{volume}{47}}, \bibinfo{pages}{16124} (\bibinfo{year}{1993}).


\bibitem[{\citenamefont{Haverkort et~al.}(2006)}]{haverkort2006:PRL-97-176405}
\bibinfo{author}{\bibfnamefont{M.~W.} \bibnamefont{Haverkort}}
  \bibnamefont{\emph{et~al.}}, \bibinfo{journal}{Phys. Rev. Lett.}
  \textbf{\bibinfo{volume}{97}}, \bibinfo{eid}{176405} (\bibinfo{year}{2006}).


\bibitem[{\citenamefont{Burnus et~al.}(2006)}]{burnus2006:PRB-74-245111}
\bibinfo{author}{\bibfnamefont{T.}~\bibnamefont{Burnus}}
  \bibnamefont{\emph{et~al.}}, \bibinfo{journal}{Phys. Rev. B}
  \textbf{\bibinfo{volume}{74}}, \bibinfo{eid}{245111} (\bibinfo{year}{2006}).


\bibitem[{\citenamefont{Csiszar et~al.}(2005)}]{csiszar2005:PRL-950187205}
\bibinfo{author}{\bibfnamefont{S.~I.} \bibnamefont{Csiszar}}
  \bibnamefont{\emph{et~al.}}, \bibinfo{journal}{Phys. Rev. Lett.}
  \textbf{\bibinfo{volume}{95}}, \bibinfo{eid}{187205} (\bibinfo{year}{2005}).


\bibitem[{\citenamefont{Mitra et~al.}(2003)}]{Mitra2003:PRB.67.092404}
\bibinfo{author}{\bibfnamefont{C.}~\bibnamefont{Mitra}}
  \bibnamefont{\emph{et~al.}}, \bibinfo{journal}{Phys. Rev. B}
  \textbf{\bibinfo{volume}{67}}, \bibinfo{pages}{092404}
  (\bibinfo{year}{2003}).


\bibitem[{CIp()}]{CIparameters}
\bibinfo{note}{Co$^{2+}$ parameters [eV]: $U_{dd}$=6.5,
$U_{cd}$=8.2, $\Delta$=6.5, $10Dq$= 0.525, $\Delta_{e_{g}}$=0.18,
$\Delta_{t_{2g}}$=0.03, $pd\sigma_{x,y}$=--1.43,
$pd\sigma_{z}$=--0.96; Co$^{3+}$: $U_{dd}$=5.5, $U_{cd}$=7.0,
$\Delta$=3.5, $10Dq$=0.785, $\Delta_{e_{g}}$= 0.24,
$\Delta_{t_{2g}}$=0.07, $pd\sigma_{x,y}$=--1.61,
$pd\sigma_{z}$=--1.16.}


\bibitem[{\citenamefont{Cwik}(2007)}]{Cwik2007}
\bibinfo{author}{\bibfnamefont{M.}~\bibnamefont{Cwik}}, Ph.D. thesis,
  \bibinfo{school}{Universit\"at zu K\"oln} (\bibinfo{year}{2007}).


\bibitem[{\citenamefont{Harrison}(1989)}]{Harrison1989}
\bibinfo{author}{\bibfnamefont{W.~A.} \bibnamefont{Harrison}},
  \emph{\bibinfo{title}{Electronic Structure and the Properties of Solids}}
  (\bibinfo{publisher}{Dover}, \bibinfo{address}{New York},
  \bibinfo{year}{1989}).


\bibitem[{\citenamefont{Blaha et~al.}(2001)}]{WIEN2k}
\bibinfo{author}{\bibfnamefont{P.}~\bibnamefont{Blaha}}
  \bibnamefont{\emph{et~al.}}, \bibinfo{title}{\emph{WIEN2k} package, http://www.wien2k.at}.


\bibitem[{\citenamefont{Anisimov et~al.}(1993)}]{Anisimov1993:PhysRevB-48-16929}
\bibinfo{author}{\bibfnamefont{V.~I.} \bibnamefont{Anisimov}}
  \bibnamefont{\emph{et~al.}}, \bibinfo{journal}{Phys. Rev. B}
  \textbf{\bibinfo{volume}{48}}, \bibinfo{pages}{16929} (\bibinfo{year}{1993}).


\bibitem[{\citenamefont{Saitoh et~al.}(1993)}]{Saitoh:PhysRevB-55-4257}
\bibinfo{author}{\bibfnamefont{T.} \bibnamefont{Saitoh}}
  \bibnamefont{\emph{et~al.}}, \bibinfo{journal}{Phys. Rev. B}
  \textbf{\bibinfo{volume}{55}}, \bibinfo{pages}{4257} (\bibinfo{year}{1997}).


\end{thebibliography}

\end{document}